\documentclass[aps,prl,preprint,groupedaddress]{revtex4-1}
\usepackage[top=1in, bottom=1.25in, left=1.25in, right=1.25in]{geometry}
\usepackage[ngerman,english]{babel}
\usepackage[utf8]{inputenc}
\usepackage[T1]{fontenc}
\usepackage{textcomp}
\usepackage{nicefrac}
\usepackage{units}
\usepackage{graphicx}
\graphicspath{{gfx/}}
\usepackage{amsmath}

\begin{document}
\title{Direct Measurement of Focusing Fields in Active Plasma Lenses} 

\author{J.-H. Röckemann$^1$}
\author{L. Schaper$^1$}
\author{S. K. Barber$^2$}
\author{N. A. Bobrova$^3$}
\author{G. Boyle$^1$}
\author{S. Bulanov$^2$}
\author{N. Delbos$^4$}
\author{K. Floettmann$^1$}
\author{G. Kube$^1$}
\author{W. Lauth$^5$}
\author{W. P. Leemans$^2$}
\author{V. Libov$^{1,6}$}
\author{A. Maier$^4$}
\author{M. Meisel$^6$}
\author{P. Messner$^4$}
\author{P. V. Sasorov$^3$}
\author{C. B. Schroeder$^2$}
\author{J. van Tilborg$^2$}
\author{S. Wesch$^1$}
\author{J. Osterhoff$^{1,}$\footnote{correspoding author: jens.osterhoff@desy.de}}
\affiliation{$^1$Deutsches Elektronen-Synchrotron DESY, Notkestraße 85, 22607 Hamburg, Germany\\
$^2$Lawrence Berkeley National Laboratory, University of California, Berkeley, California 94720, USA\\
$^3$Keldysh Institute for Applied Mathematics, Moscow 125047, Russia\\
$^4$Center for Free-Electron Laser Science and Department of Physics, Luruper Chaussee 149, 22761 Hamburg, Germany\\
$^5$Johannes Gutenberg-Universität Mainz, Saarstraße 21, 55122 Mainz, Germany\\
$^6$Universität Hamburg, Mittelweg 177, 20148 Hamburg, Germany
}

\date{\today}

\begin{abstract}
Active plasma lenses have the potential to enable broad-ranging applications of plasma-based accelerators owing to their compact design and radially symmetric kT/m-level focusing fields, facilitating beam-quality preservation and compact beam transport. We report on the direct measurement of magnetic field gradients in active plasma lenses and demonstrate their impact on the emittance of a charged particle beam. This is made possible by the use of a well-characterized electron beam with 1.4\,mm\,mrad normalized emittance from a conventional accelerator. Field gradients of up to 823\,T/m are investigated. The observed emittance evolution is supported by numerical simulations, which suggest the potential for conservation of the core beam emittance in such a plasma lens setup.
\end{abstract}
\maketitle

\section{Introduction}
Laser wakefield accelerators (LWFAs) allow for the generation of extreme electric fields on the order of 100\,GV/m for charged particle acceleration and can deliver beams of sub-\textmu m normalized emittance \cite{Barber2017, Weingartner2012}, multi-kA peak currents \cite{geddes2004}, and femtosecond pulse duration \cite{LundhNature,HZDRNatComm, buck2011}. LWFAs have shown the capability to produce multi-GeV electron beams in cm-scale structures \cite{kim2013,Wang2013,leemans2014}. Their application to drive compact sources of coherent X-ray beams \cite{maier2012,huang2012} and incoherent MeV photons \cite{chen2013mev}, ultra-fast electron diffraction experiments \cite{Esarey2001,Hartemann2007}, and high-energy particle colliders \cite{schroeder2010} has been proposed and studied \cite{glinec2005,powers2014quasi}. For all these applications small beam emittances are critical. Indeed, beams from plasma accelerators are susceptible to chromatic emittance growth in the drift following the acceleration section \cite{Floettmann2003,mehrling2012}. Thus, beam capturing within a few centimeters after the plasma exit is crucial for emittance preservation.\\
In this context, conventional focusing optics face problems: Solenoids suffer from large chromaticity and weak focusing for relativistic beams owing to their $1/\gamma^2$-scaling of the focusing strength, with the relativistic Lorentz factor $\gamma$. The more favorable $1/\gamma$-scaling in combination with high field gradients ($\sim$500\,T/m for permanent magnets) of quadrupoles is put into perspective when considering that two quadrupoles need to be combined to achieve focusing in both transverse planes. Hence, quadrupoles, which are inherently defocusing in one plane, increase chromatic emittance growth in this plane dramatically \cite{Lindstrom2016}.\\
Active plasma lenses \cite{Forsyth1965} (APLs) potentially offer an elegant solution with their compact size, azimuthally symmetric focusing, and high magnetic field gradients, which have been shown to exceed 3\,kT/m \cite{vanTilborg2015}. Recent studies indicate that nonuniform current densities may form inside discharge capillary based APLs	 \cite{bobrova2001simulations,spence2000investigation,butler2002guiding,mcguffey2009guiding,gonsalves2016demonstration}, leading to nonlinear magnetic field gradients and, subsequently, emittance deterioration \cite{vanTilborg2017,pompili2017experimental}. In this work we report on a first direct measurement of the magnetic field distribution inside an APL and complement these results by experimentally detecting its influence on the emittance of a stable, well-characterized electron beam from a conventional accelerator. These studies are supported by simulations and show the potential for emittance preservation.\\
Active plasma lenses for electron beams typically consist of a gas-filled capillary with a circular cross-section of mm-scale diameter and cm-scale length machined into glass or sapphire. A multi-kV discharge voltage is applied to the capillary ends, leading to breakdown of the gas. Subsequently, a current is driven along the generated plasma column forming an azimuthal magnetic field. In the following, we assume a azimuthally symmetric current distribution $J(r)$, with $r$ denoting the radial position. Ampere's law provides the cylindrically symmetric magnetic field
\begin{equation}
B_{\phi} (r) \cdot r = \mu_0 \int_0^r J(r') r' dr',
\label{ampereslaw}
\end{equation}
for $r\,<\,R$, with $R$ being the capillary radius and $\mu_0$ the vacuum permeability. The magnetic field distribution becomes $B_{\phi, \text{ideal}} (r) = \mu_0 I_0 r/(2 \pi R^2)$ in case of a uniform current distribution $J = I_0/(\pi R^2)$, with $I_0$ being the total current. Differentiating this expression yields the ideal magnetic field gradient
\begin{equation}
g_{\text{ideal}} = \mu_0 I_0 / (2 \pi R^2).
\label{uniformAPL}
\end{equation}

\section{Nonlinear model of active plasma lenses}
\begin{figure}
\centering
\includegraphics[width=\columnwidth]{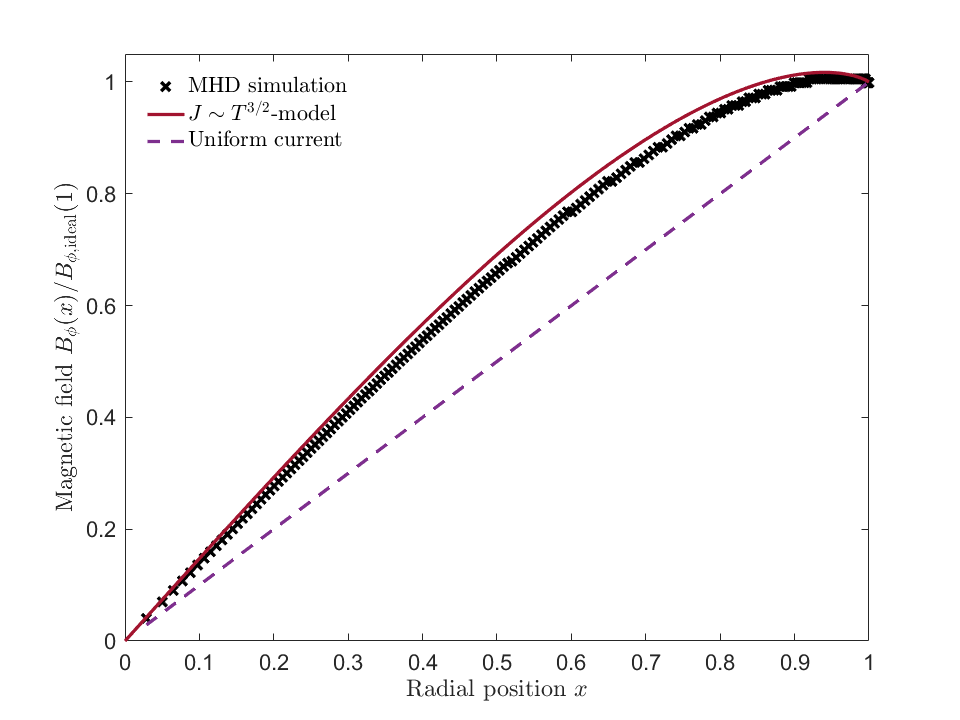}
\caption{MHD simulation results for a $R = 0.5$\,mm gas column with $I_0 = 364$\,A. The $J \sim T^{3/2}$-model is of the form of Eq.\,\eqref{a4APL}.}
\label{MHDFieldPlot}
\end{figure}

In general, Eq.\,\eqref{uniformAPL} does not hold since the current distribution homogeneity $J(r)$ is generally not uniform. A transverse temperature gradient forms due to cooling of the plasma at the capillary wall leading to a radially changing Ohmic resistance, a nonuniform current distribution, and a nonlinear magnetic field gradient. Fig.\,\ref{MHDFieldPlot} shows the result of a one-dimensional magnetohydrodynamic (MHD) simulation of a capillary of $R=0.5$\,mm radius filled with hydrogen of $n_0 = 10^{17} cm^{-3}$ molecular density traversed by a current of $I_0=364$\,A assuming a fixed electron temperature at the wall interface of $T^{*} = 0.5$\,eV. The radial position is normalized to $R$, the magnetic field to $B_{\text{ideal}}$. Cases with $I_0=188$\,A, and 740\,A have also been simulated. The MHD modeling shows that for the currents used, the fraction of ionized hydrogen was well above 80\%.\\
An analytic model for the current distribution in a plasma lens was introduced in \cite{vanTilborg2017}. It is based on the Spitzer collisional model of plasma, in which the conductivity $\sigma$ depends on the plasma density $n_e$ and electron temperature $T_e$ via
\begin{equation}
\sigma = \frac{32 \epsilon_{0}^2}{ln \Lambda} \cdot \frac{(k_{B} T_{e})^{3/2}}{e^2 m_{e}^{1/2}},
\label{plasmaconductivity}
\end{equation}
with $\lambda_{D} = \sqrt{\epsilon_0 k_B T_e / n_e e^2}$, $\Lambda = n_e \lambda_{D}^3$, $k_B$ the Boltzmann constant, $\epsilon_0$ the vacuum permittivity, $e$ the electron charge, and $m_{e}$ the electron mass. The scaling of $\sigma$ is dominated by $T_e$ since $n_e$ appears only in the logarithm of $\Lambda$. Thus, the current density is dominated by the temperature $J(r) = \sigma E \sim T_e^{3/2}$. Following the work of \cite{bobrova2001simulations,vanTilborg2017}, the temperature distribution satisfies the heat flow equation
\begin{equation}
\frac{1}{x}\frac{d}{dx} \left( x \frac{du}{dx} \right) = -u^{3/7},
\label{heatflow}
\end{equation}
in which $u^{2/7} = T_e / A$ with $A = (7 \sigma_0 R^2 E^2 / 2 \kappa_0 )^{1/2}$, $x=r/R$, and the thermal and electric conductivities were assumed to scale with $\kappa = \kappa_0 T_e^{5/2}$ and $\sigma = \sigma_0 T_e^{3/2}$, respectively. The boundary conditions satisfy $d T_e (x=0) / dx =0$ and $T_e(x=1) = T^{*}$ with $T^{*}$ the temperature at the wall. The current distribution can be expressed as
\begin{equation}
J(r) = \frac{I_0 u(r)^{3/7}}{2 \pi R^2 m_I},
\label{currentdist}
\end{equation}
with $m_I = \int_0^1 u^{3/7} x dx$. The central region $x<1$ can be written as
\begin{equation}
J (r) = \frac{I_0}{\pi R^2} \left( \frac{u(0)^{3/7}}{2 m_I} \right) \left[ 1 - \frac{3}{28} u(0)^{-4/7} x^2 - \frac{15}{3136}u(0)^{-8/7} x^4 \right],
\label{tayloredcurr}
\end{equation}
and
\begin{equation}
B_{\phi} (x) = \frac{\mu_0 I_0}{2 \pi R} \cdot \frac{u(0)^{3/7}}{2m_I} \cdot x \cdot \left[ 1 - \frac{3}{56}u(0)^{-4/7} x^2 - \frac{5}{3136}u(0)^{-8/7} x^4 \right].
\label{a4APL}
\end{equation}
An important figure of merit for the linearity of an APL is its core linear magnetic field slope in comparison to the ideal magnetic field slope, defined as $\Delta\,g\,=\,g_{\text{core}}/g_{\text{ideal}} = \frac{u(0)^{3/7}}{2m_I}$. The $\Delta\,g$-factor for the $J \sim T^{3/2}$-model in Fig.\,\ref{MHDFieldPlot} is $\Delta\,g = 1.48$. This corresponds to a cold wall boundary condition. The corresponding gradients are given in Tab.\,\ref{table}.
\section{Experimental setup}
\begin{figure}
\centering
\includegraphics[width=\columnwidth]{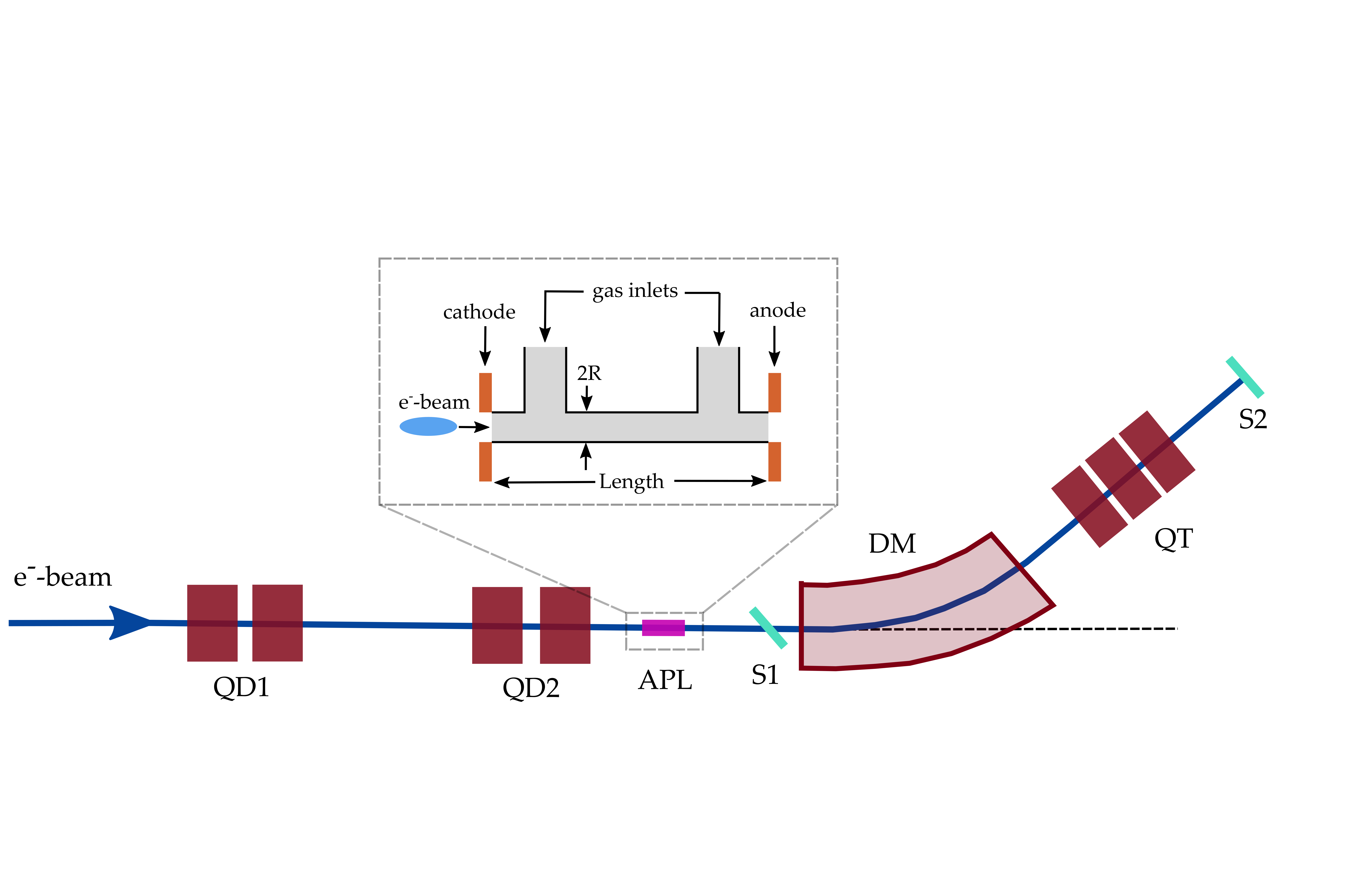}
\caption{Schematic of the accelerator beamline at MaMi-B. QD1: first quadrupole duplet; QD2: second quadrupole duplet; APL: active plasma lens; S1: screen used in the offset measurements; DM: dipole magnet; QT: quadrupole triplet used in the emittance measurements; S2: screen used in the emittance measurements.}
\label{MaMiBeamline}
\end{figure}

The APL in this experiment consisted of a 7\,mm long capillary of $R = 0.5$\,mm machined into a sapphire block. A continuous flow of hydrogen was supplied to the capillary at 4\,mbar backing pressure through two inlets of $R= 0.75$\,mm diameter situated 1.5\,mm from the capillary ends leading to a molecular density of $n_0 = 10^{17} cm^{-3}$ inside the capillary. Copper electrodes on both sides connected a pulse-forming network \cite{donald1956pulse} to the gas volume. A discharge voltage of 9\,kV - 20\,kV was applied which resulted in stable flat-top currents of 188\,A - 740\,A arising 100\,ns after the discharge trigger for a duration of 240\,ns. The electron beam traversed the APL 100\,ns after the current plateaued. A schematic drawing of the APL insise the experimental setup is given in Fig.\,\ref{MaMiBeamline}.\\
%\begin{figure}
%\centering
%\includegraphics[width=0.6\columnwidth]{SketchAPL}
%\caption{Schematic of the active plasma lens setup. The electrodes fit into the sapphire slab forming a gap of 7\,mm.}
%\label{APLSketch}
%\end{figure}
A race-track Microtron at the University of Mainz, the Mainz Microtron B (MaMi-B), was used for probing the magnetic field of the APL. MaMi-B was operated in a mode in which it delivered 10\,ns long bunches with an average current of 100\,\textmu A, an energy of 855\,MeV, and a normalized vertical emittance of $\epsilon_{\text{i}} = 1.37 \pm 0.01$\,mm\,mrad.
\section{Experimental results}
\begin{figure}
\includegraphics[width=\columnwidth]{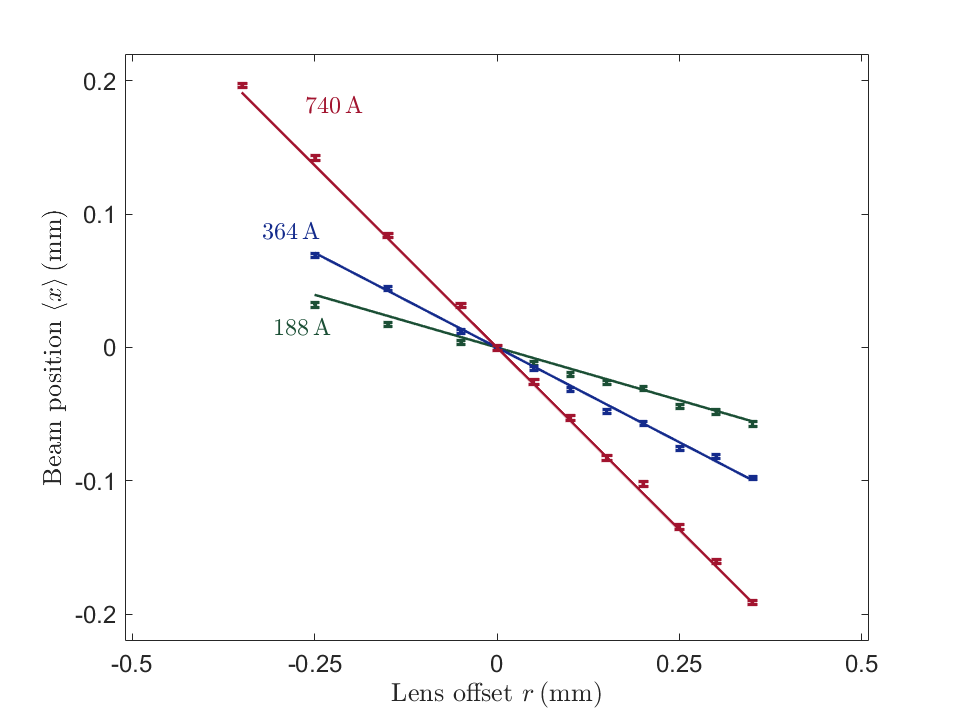}
\caption{Results from offset scan using $I_0 = 188$\,A, 364\,A and 740\,A of total current. The lines are linear fits to the data. The error bars include statistical fluctuations and underline the stability of the APL setup.}
\label{7mmGradients}
\end{figure}

\begin{figure}
\includegraphics[width=\columnwidth]{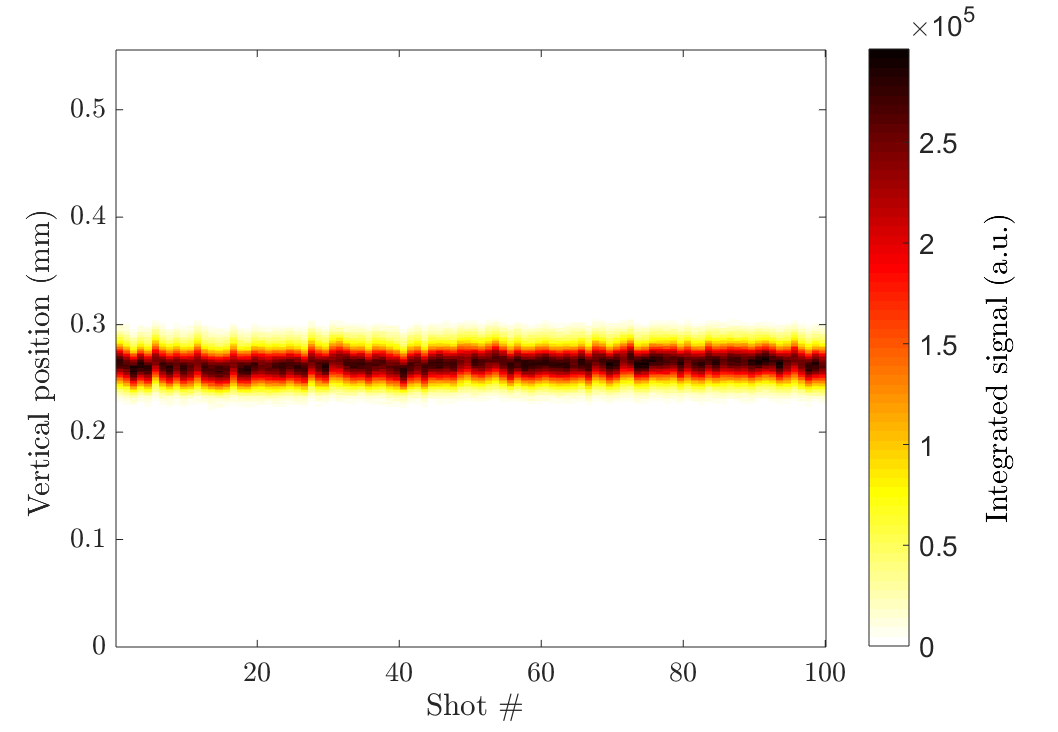}
\caption{Camera signal projected onto the vertical axis for 100 consecutive shots at 350\,\textmu m APL offset. The excellent stability of MaMi was not measurably affected by the APL, which is indicated by the small shot-to-shot fluctuations of the signal.}
\label{ShotStab}
\end{figure}

\begin{figure}
\includegraphics[width=\columnwidth]{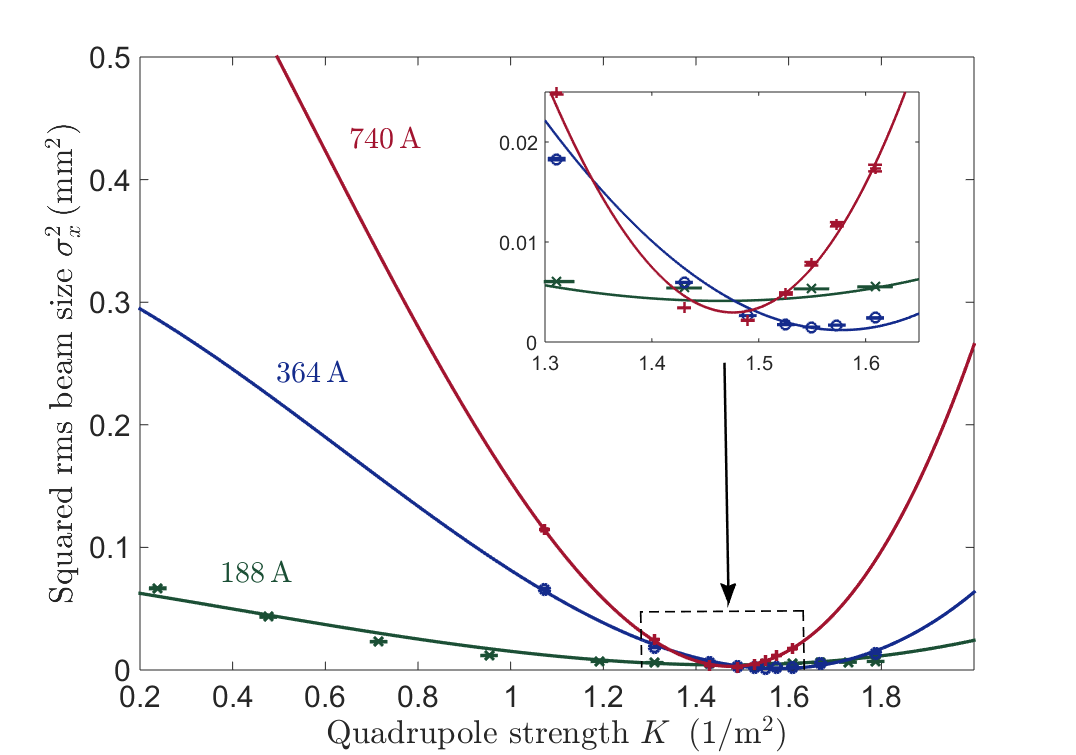}
\caption{Quadrupole scan results for the APL operated at 188\,A, 364\,A, and 740\,A of total current. Error bars for beam size measurements are included. The fitted emittances can be found in Tab.\,\ref{table}.}
\label{Emittance7mm}
\end{figure}

\begin{figure}
\includegraphics[width=\columnwidth]{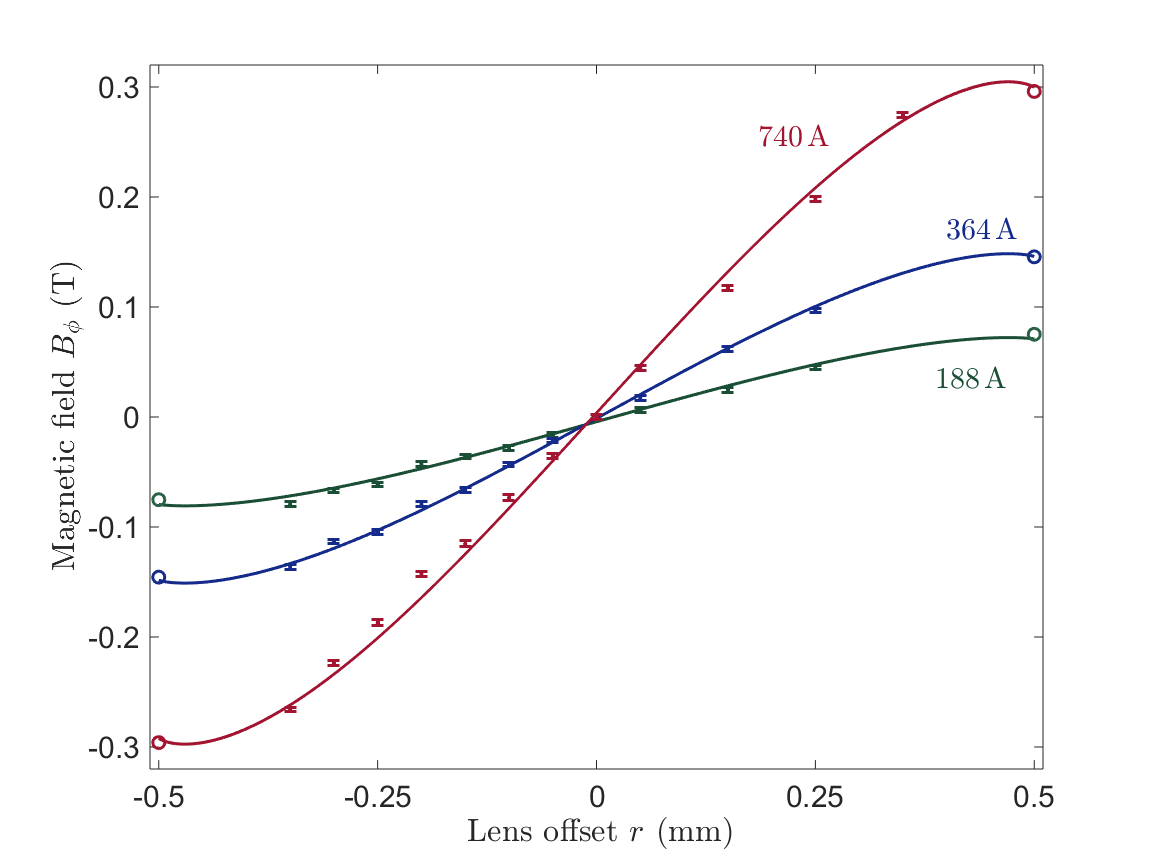}
\caption{Magnetic field strength in an $L=7.5$\,mm APL derived from the offset scan data in Fig.\,\ref{7mmGradients}, for $r = \pm 0.5$\,mm, obtained from measurements of $R$ and $I_0$ (circles). The sizes of the circular markers represent the rms error of the data points. The lines show the predicted behavior from the $J \sim T^{3/2}$-model.}
\label{7mmBField}
\end{figure}

Direct measurements of the APL magnetic field gradients were performed by introducing a transverse offset of the APL with respect to the electron beam position, thus introducing a dipole kick to the beam. The centroid shifts and beam parameters of the MaMi-B beam were measured d = 25.3cm downstream of the APL at screen S1 and averaged over 100 shots per offset position. The beam was focused into the APL in order to probe over the largest portion of the radius possible without beam clipping. Its dimensions at the capillary entrance were determined by backtracking the beam parameters based on the measurements at S1 without the plasma in its path. The beam size was calculated to be 80\,\textmu m rms in both planes. The offset was increased until clipping and charge loss of the beam became evident which resulted in a maximum offset of 350\,\textmu m. The resulting centroid shifts can be seen in Fig.\,\ref{7mmGradients} and are found to be linearly depending on the offset. The formation of fringe fields in APLswas discussed in \cite{Bagdasorov2017}. Their influence on the emittance of a passing MaMi-type beam was simulated in ASTRA \cite{FloettmannASTRA} and found to be negligible on the sub-percent level. The longitudinal current ramp in the fringe fields was modeled after $I(z_{\text{edge}}) = I_0/(1+\exp(4 z_{\text{edge}}/\sigma_{\text{ramp}}))$, where $z_{\text{edge}}$ is the distance from the capillary end and $\sigma_{ramp}$ is the ramp taper parameter, as commonly used in conventional magnet optics. Owing to the fringe fields, the effective magnetic length $L = L_{\text{capillary}} + 2\cdot L_{\text{fringe}}$ of the APL extends beyond the sapphire capillary itself. So the beam offset $ \Delta \langle x \rangle$ is dependent on the lens offset $r$ and effective length $L$ through the magnetic field
\begin{equation}
\Delta \langle x \rangle = \frac{q \cdot d}{p}\int_{0}^{L} B_{\phi}(r) z dz,
\label{offset}
\end{equation}
in which $p$ is the particle momentum, $q$ its charge. To account for the additional uncertainty owing to the fringe field, the data in Fig.\,\ref{7mmGradients} was fitted for the range of $L_{\text{fringe}} \leq 0.5$\,mm (which is well above the length found in \cite{Bagdasorov2017}). The derived core gradients $g_{\text{core}}$ for $L_{\text{fringe}}= 0.25$\,mm including the systematic uncertainty for $L_{\text{fringe}} \leq 0.5$\,mm can be found in Tab.\,\ref{table}. The obtained magnetic field gradients are higher than Eq.\,\eqref{uniformAPL} would predict from the measured discharge currents. They are, however, in good agreement with a $J \sim T^{3/2}$-model assuming a cold wall boundary condition with $\Delta\,g \sim 1.48$. It is noteworthy that the relative center-of-mass jitter of the MaMi-B beam was not affected by the APL even for the extreme case of 350\,\textmu m offset (cf. Fig.\,\ref{ShotStab}). This means the magnetic field in the APL was highly reproducible, which may also be seen in the small error bars of the measured beam position in Fig.\,\ref{7mmGradients}.\\
A complementary way of probing the linearity of the magnetic field in the APL is measuring the emittance change of an electron beam after passage through the APL. Quadrupole scans were performed for different plasma lens settings in order to detect emittance change due to nonlinear field gradients. The currents used in the experiment were 188, 364, and 740\,A. The current amplitude had a jitter of 1.5\,A rms in each case. This measurement technique requires the beamline upstream of the quadrupoles used for the scan to be stable. The here reported APL stability greatly facilitated these emittance measurements and is reflected in the relatively low rms beam size variation during the scans of $ < 5\%$ (100 shots were averaged per setting). In order to probe for nonlinearities over a large fraction of the capillary diameter an rms beam size of $\sigma_{y} =  154^{+5}_{-15}$\,\textmu m vertically and $\sigma_{x} = 151^{+2}_{-12}$\,\textmu m horizontally and a small divergence $\sigma_{x\prime,y\prime} < 0.1$\,mrad at the APL entrance was used for the emittance scans. The results of three quadrupole scans are shown in Tab.\,\ref{table}. The measured beam sizes including fits can be found in Fig.\,\ref{Emittance7mm}. The optical system had a resolution of 20\,\textmu m which was well suited for the 45\,\textmu m of minimal beam size used in the scans.\\
%Since emittance change is an additive effect, the introduced emittance by the APL $\epsilon_{\text{APL}}$ is derived from $\epsilon_{\text{f}}^2 = \epsilon_{\text{APL}}^2 + \epsilon_{\text{i}}^2$.
At first glance, the offset measurements in Fig.\,\ref{7mmGradients} seem to imply, that field nonlinearity is not the root cause of the observed emittance degradation since the linear fits show excellent agreement with the data. Utilizing measurements of $R$ and $I_0$ allows to derive the magnetic field strength at the wall through Eq.\,\eqref{ampereslaw}, showing that the linear behavior measured over the central 70\% of the capillary diameter fails to describe the magnetic field for the entire capillary width. Since Eq.\,\eqref{ampereslaw} is always fulfilled at the wall and $R$ and $I_0$ were measured with high precision, the derived magnetic field values have small errors of $< \pm 3 \%$. The $J \sim T^{3/2}$-model is in good agreement with all of the measurements including the detected emittance growth (see next section and Tab.\,\ref{table}). Fig.\,\ref{7mmBField} shows the predicted behavior from the $J \sim T^{3/2}$-model assuming a cold wall boundary condition on top of the derived magnetic field values from the offset scan in Fig.\,\ref{7mmGradients} and as additional data points the magnetic field at the wall ($r= \pm R$) from Eq.\,\eqref{ampereslaw}.\\
Other mechanisms for the emittance degradation such as self-wakefields and collisions fail to describe the observed dependence on total current which can readily be explained by a non linear field model. The driving of a self-wake can be neglected because of the low peak current used in the beam \citep{lindstrom2018analytic}. The emittance growth due to collisions can be estimated for: a) multiple scattering in neutral background gas \citep{Reiser}, and b) transport in plasma \citep{Montague, Humphries} including the stopping power effects of collisions with free, bound and screened electrons, and Bremsstrahlung \citep{Humphries, Touati}. For the parameters relevant to this work, the normalized emittance growth due to scattering is estimated to be $< 0.05$\,mm\,mrad. Owing to the small energy spread of MaMi ($\sim 10^{-5}$), chromatic effects were not relevant. The chromaticity introduced by the beam-plasma interaction was measured in the emittance measurements due to the dispersion introduced by the dipole in between the APL and the QMs used for the scans (cf. Fig.\,\ref{MaMiBeamline}). No broadening of the energy spread was observed confirming the non-existence of self-wakefields.\\
\section{Simulation results}
\begin{figure}
\includegraphics[width=\columnwidth]{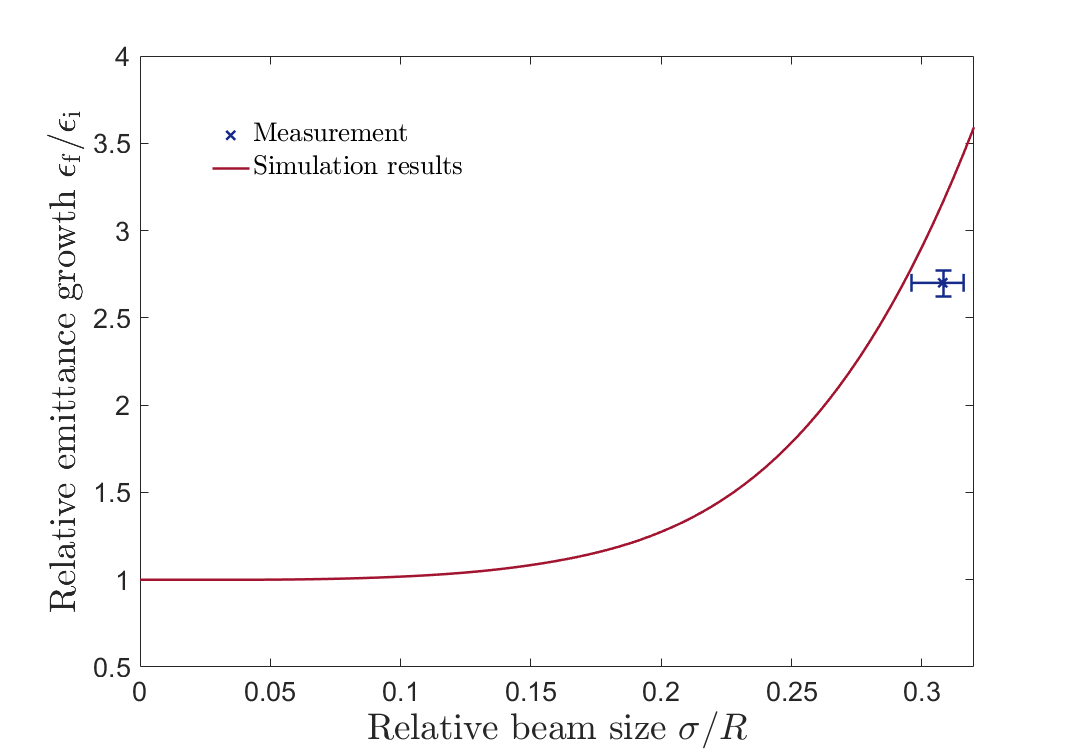}
\caption{Particle tracking simulation results for relative emittance degradation of a MaMi-like beam in dependence of incoming rms beam size for an APL with $R=0.5$\,mm and $I_0 = 364$\,A. The measured emittance degradation for this setup is also shown.}
\label{EmittanceASTRA}
\end{figure}

The emittance growth in an $L = 7.5$\,mm long APL was simulated with the particle tracking code ASTRA. The field was modeled to be of the form given by the $J \sim T^{3/2}$-model. Transversally Gaussian shaped beams with rms beam size of $\sigma_{x,y} = 154$\,\textmu m were assumed for the simulation. The measured emittance growth and the simulation results in Tab.\,\ref{table} are in excellent agreement. This supports the conjecture that the $J \sim T^{3/2}$-model is a good approximation for the field behavior in the APL. Fig.\,\ref{EmittanceASTRA} shows particle tracking simulation results for different incoming rms beam sizes traversing the magnetic field distributions obtained from the $J\sim T^{3/2}$-model and $L=7.5$\,mm and 364\,A. They suggest a smaller beam size in the APL than used here is favorable to minimize emittance growth. The emittance growth is highly dependent on the incoming beam size and can be effectively eliminated for beams with $\sigma < 75$\,\textmu m on the 1\,mm\,mrad  normalized emittance scale according to these numerical results. The data point shown in Fig.\,\ref{EmittanceASTRA} corresponds to the measuremed emittance from Fig.\,\ref{Emittance7mm} for 364\,A.\\
\begin{table}
  \caption{Comparison of measured and simulated gradients and emittances. The measured gradients are for an effective length of $L=7.5$\,mm. Additionally, systematic uncertainties arising from the fringe fields are given. The emittance was simulated for $\sigma = 154$\,\textmu m and a field in the form of the $J \sim T^{3/2}$-model (cf. Fig.\,\ref{MHDFieldPlot}).}
  \label{table}\centering
	\begin{tabular}{c|c|c|c|c|c|c}
         $I_0$ & \multicolumn{3}{c|}{$g_{\text{core}}$\,(T/m)} & \multicolumn{2}{c}{$\epsilon_{\text{f}}$\,(mm\,mrad)}  \\
		\,(A) & Meas. & Fringe & $J \sim T^{3/2} $ & Meas. & Sim.
        \\ \hline
      	188 & 238\,$\pm$\,9 & $\pm$\,17 & 223 & 2.2\,$\pm$\,0.1 & 2.5 \\
        364 & 428\,$\pm$\,6 & $\pm$\,30 & 431 & 3.7\,$\pm$\,0.1 & 4.3 \\
      	740 & 823\,$\pm$\,8 & $\pm$\,59 & 876 & 8.2\,$\pm$\,0.1 & 8.4 \\
	\end{tabular}
\end{table}
%\begin{figure}
%\includegraphics[width=\columnwidth]{OffsetScan_JTcoldwall_550muFringe}
%\caption{Results from offset scan using 188\,A, 364\,A and 740\,A with the magnetic field at the capillary wall ($r=R$) calculated from the currents. Ideal case gradients are shown in dashed lines, the polynomial fits are of the form of Eq.\,\eqref{a4APL} with a cold wall boundary condition. Error bars are included for both offset scan and current measurements.}
%\label{7mmGradients}
%\end{figure}
\section{Conclusion}
In summary, magnetic field gradients of a 1-mm diameter active plasma lens and the emittance change of a beam passing such a lens have been measured directly using the conventional accelerator Mainz Microtron. We observed excellent gradient stability. The measured gradient increase of $\Delta g \simeq 1.5$ showed a behavior predicted for a cold wall boundary condition $J\sim T^{3/2}$-model. The measured emittance change of a passing electron beam agrees with predictions made by magnetohydrodynamics simulations and particle tracking simulations using the measured gradient enhancement as input parameter. Furthermore, simulations suggest that using beams of an rms size smaller than 20\% of the APL radius leads to emittance preservation on the mm\,mrad-level. Future studies will focus on mitigating emittance degradation further by manipulating the current density behavior in the APL by using different gas species and optimizing radii and current profiles.
\section{Acknowledgements}
We acknowledge the support through the Helmholtz Virtual Institute VH-VI-503, the Helmholtz Matter and Technologies Accelerator Research and Development program, the Helmholtz IuVF ZT-0009 program, and the U.S. Dept. of Energy under contract No. DE-AC02-05CH11231.
\clearpage
\bibliography{iopbib}

\begin{thebibliography}{10}

\bibitem{Barber2017}
S.~Barber, {\it et~al.\/}, {\it Phys. Rev. Lett.\/} {\bf 119}, 104801 (2017).

\bibitem{Weingartner2012}
R.~Weingartner, {\it et~al.\/}, {\it Physical Review Accelerators and Beams\/}
  {\bf 15}, 111302 (2012).

\bibitem{geddes2004}
C.~Geddes, C.~Toth, J.~van Tilborg, E.~Esarey, {\it et~al.\/}, {\it Nature\/}
  {\bf 431}, 538 (2004).

\bibitem{LundhNature}
O.~Lundh, {\it et~al.\/}, {\it Nature Physics\/} {\bf 7}, 219 (2011).

\bibitem{HZDRNatComm}
K.~Zeil, {\it et~al.\/}, {\it Nature Communications\/} {\bf 3}, 219 (2012).

\bibitem{buck2011}
A.~Buck, {\it et~al.\/}, {\it Nature Physics\/} {\bf 7}, 543 (2011).

\bibitem{kim2013}
H.~T. Kim, {\it et~al.\/}, {\it Physical Review Letters\/} {\bf 111}, 165002
  (2013).

\bibitem{Wang2013}
X.~Wang, {\it et~al.\/}, {\it Nature Communications\/} {\bf 4} (2013).

\bibitem{leemans2014}
W.~Leemans, {\it et~al.\/}, {\it Physical Review Letters\/} {\bf 113}, 245002
  (2014).

\bibitem{maier2012}
A.~Maier, {\it et~al.\/}, {\it Physical Review X\/} {\bf 2}, 031019 (2012).

\bibitem{huang2012}
Z.~Huang, Y.~Ding, C.~B. Schroeder, {\it Physical Review Letters\/} {\bf 109},
  204801 (2012).

\bibitem{chen2013mev}
S.~Chen, {\it et~al.\/}, {\it Physical Review letters\/} {\bf 110}, 155003
  (2013).

\bibitem{Esarey2001}
P.~Catravas, {\it et~al.\/}, {\it Meas. Sci. Technol.\/} {\bf 12}, 1828 (2001).

\bibitem{Hartemann2007}
F.~Hartemann, {\it et~al.\/}, {\it IEEE Transactions on Nuclear Science\/} {\bf
  10}, 011301 (2007).

\bibitem{schroeder2010}
C.~Schroeder, E.~Esarey, C.~Geddes, C.~Benedetti, W.~Leemans, {\it Physical
  Review Special Topics-Accelerators and Beams\/} {\bf 13}, 101301 (2010).

\bibitem{glinec2005}
Y.~Glinec, {\it et~al.\/}, {\it Physical Review Letters\/} {\bf 94}, 025003
  (2005).

\bibitem{powers2014quasi}
N.~D. Powers, {\it et~al.\/}, {\it Nature Photonics\/} {\bf 8}, 28 (2014).

\bibitem{Floettmann2003}
K.~Floettmann, {\it Physical Review Special Topics-Accelerators and Beams\/}
  {\bf 6}, 034202 (2003).

\bibitem{mehrling2012}
T.~Mehrling, J.~Grebenyuk, F.~Tsung, K.~Floettmann, J.~Osterhoff, {\it Physical
  Review Special Topics-Accelerators and Beams\/} {\bf 15}, 111303 (2012).

\bibitem{Lindstrom2016}
C.~Lindstr{\o}m, E.~Adli, {\it Physical Review Special Topics-Accelerators and
  Beams\/} {\bf 19}, 071002 (2016).

\bibitem{Forsyth1965}
E.~Forsyth, {\it et~al.\/}, {\it Meas. Sci. Technol.\/} {\bf 12}, 872 (1965).

\bibitem{vanTilborg2015}
J.~van Tilborg, {\it et~al.\/}, {\it Physical Review Letters\/} {\bf 115},
  184802 (2015).

\bibitem{bobrova2001simulations}
N.~Bobrova, {\it et~al.\/}, {\it Physical Review E\/} {\bf 65}, 016407 (2001).

\bibitem{spence2000investigation}
D.~Spence, S.~M. Hooker, {\it Physical Review E\/} {\bf 63}, 015401 (2000).

\bibitem{butler2002guiding}
A.~Butler, D.~Spence, S.~M. Hooker, {\it Physical Review Letters\/} {\bf 89},
  185003 (2002).

\bibitem{mcguffey2009guiding}
C.~McGuffey, {\it et~al.\/}, {\it Physics of Plasmas\/} {\bf 16}, 113105
  (2009).

\bibitem{gonsalves2016demonstration}
A.~Gonsalves, {\it et~al.\/}, {\it Journal of Applied Physics\/} {\bf 119},
  033302 (2016).

\bibitem{vanTilborg2017}
J.~van Tilborg, {\it et~al.\/}, {\it Physical Review Accelerators and Beams\/}
  {\bf 20}, 032803 (2017).

\bibitem{pompili2017experimental}
R.~Pompili, {\it et~al.\/}, {\it Applied Physics Letters\/} {\bf 110}, 104101
  (2017).

\bibitem{donald1956pulse}
P.~G. Donald, Pulse forming network (1956). US Patent 2,769,903.

\bibitem{Bagdasorov2017}
G.~Bagdasorov, {\it et~al.\/}, {\it Physics of Plasmas\/} {\bf 24}, 083109
  (2017).

\bibitem{FloettmannASTRA}
K.~Floettmann, Astra particle tracking code. {www.desy.de/}$\sim${mpyflo}.

\bibitem{lindstrom2018analytic}
C.~A. Lindstr{\o}m, E.~Adli, {\it arXiv preprint arXiv:1802.02750\/}  (2018).

\bibitem{Reiser}
M.~Reiser, {\it Theory and design of charged particle beams\/} (John Wiley \&
  Sons, 2008).

\bibitem{Montague}
B.~Montague, W.~Schnell, {\it AIP Conference Proceedings\/} (AIP, 1985), vol.
  130, pp. 146--155.

\bibitem{Humphries}
S.~Humphries, {\it Charged particle beams\/} (Courier Corporation, 2013).

\bibitem{Touati}
M.~Touati, {\it et~al.\/}, {\it New Journal of Physics\/} {\bf 16}, 073014
  (2014).

\end{thebibliography}
\bibliographystyle{Science}
\end{document}